\documentclass[twocolumn,pra,aps,superscriptaddress,showpacs]{revtex4-1}
\usepackage{mathptmx}
\usepackage{subfigure}
\usepackage{psfrag,graphicx}
\usepackage{pict2e}
\usepackage{dcolumn}
\usepackage{latexsym}
\usepackage{epstopdf}
\usepackage{color}
\usepackage[english]{babel}
\usepackage{amsmath}
\usepackage{amssymb}
\usepackage{amsfonts}
\usepackage{bm}
\usepackage{natbib}
\usepackage{appendix}
\definecolor{mygrey}{gray}{0.35}
\definecolor{myblue}{rgb}{0.2,0.2,0.8}
\definecolor{myzard}{cmyk}{0,0,0.05,0}
\definecolor{mywhite}{rgb}{1,1,1}
\definecolor{mywhite}{rgb}{1,1,1}
\definecolor{myred}{rgb}{1,0.,0.3}
\usepackage[colorlinks=true,citecolor=myblue,linkcolor=myred]{hyperref}
\DeclareMathOperator{\Li}{Li}
\begin{document}
\title{Hidden frustrated interactions and quantum annealing in trapped-ion spin-phonon chains}

\author{Pedro Nevado}
\affiliation{Department of Physics and Astronomy, University of Sussex, Falmer, Brighton BN1 9QH, UK}

\author{Diego Porras}
\email[Corresponding author: ]{d.porras@sussex.ac.uk}
\affiliation{Department of Physics and Astronomy, University of Sussex, Falmer, Brighton BN1 9QH, UK}

\date{\today}

\begin{abstract}

We show that a trapped-ion chain interacting with an optical spin-dependent force shows strong frustration effects due to the interplay between long range interactions and the dressing by optical phases. We consider a strong spin-phonon coupling regime and predict a quantum phase diagram with different competing magnetic and structural orders. In a highly frustrated region, the system shows enhanced quantum fluctuations and entanglement, which are characteristic of spin-liquid phases. We propose and describe within a mean-field approach a quantum annealing process to induce a quasi-adiabatic evolution towards the ground state.

\end{abstract}


\maketitle

\section{Introduction}

Systems with frustrated interactions play an important role in fields from magnetism \cite{balents10nat} to molecular biology \cite{onuchic97arpc}. Frustration may lead to a complex energy landscape where finding the global energy minimum becomes a computationally demanding task.
Those systems exhibit fascinating properties, like the strong thermal or quantum fluctuations that characterize spin-liquid phases \cite{balents10nat,Ramirez94}. 
Atomic analogical quantum simulators \cite{Cirac12natphys} are an ideal test-bed to explore frustration and quantum annealing. Trapped ion setups have emerged in the last years as a powerful platform to simulate quantum magnetism \cite{Porras04prl1,Friedenauer08natphys,Schneider12rpp,Blatt12natphys}. 
Recent experiments have demonstrated the simulation of frustrated Ising models with up to 16 ions \cite{Kim10nat, Islam13sci} in linear traps, and the generation of Ising interactions with hundreds of ions in planar Penning traps \cite{britton12nat}. The technical progress in the fabrication of two-dimensional arrays of microtraps has an exciting outlook in providing us with synthetic ion crystals of different geometries that can lead to geometrical frustration \cite{Schneider12rpp,sterling14natcom}.
Zigzag ion crystal phases have been also proposed to study frustrated quantum  magnets \cite{Bermudez11prl}.

In this work we depart from schemes simulating effective spin-spin couplings \cite{Porras04prl1} and consider the limit of strong spin-phonon coupling leading to cooperative Jahn-Teller \cite{Porras12prl} or Rabi lattice models \cite{Koch13njp,Kurcz14prl,Nevado2015PRA}. We unveil the existence of hidden long range frustrated interactions leading to the competition between different magnetic and structural orders. Frustration arises by the interplay of two effects: (a) an effective long range coupling between different ions  and (b) the dressing of spin-phonon interactions by means of the optical phases of lasers. Our scheme can be easily implemented with many ions, since it only relies on a single optical force acting on the ion chain. It also allows high simulation speeds, since we consider a non-perturbative spin-phonon coupling regime.

This article is structured as follows: (i) We present our model, focusing on the case of a trapped-ion spin-phonon chain. (ii) We study the phase diagram of the model, assuming no optical phases from the lasers. In the classical limit of the problem, we show the existence of a set of (classical) phase transitions separating different magnetic/structural configurations. Separately, we calculate the quantum phase diagram by means of the Density Matrix Renormalization Group (DMRG) method \cite{White92prl,Schollwock11anp}. (iii) Next, we carry out the same two-part study for the case of non-zero optical phases. Owing to these, the system exhibits frustration. We identify the values of the parameters controlling frustration in the ion chain, and pinpoint a highly frustrated regime where quantum fluctuations and entanglement are enhanced. (iv) We propose a quantum annealing scheme that can be used in a trapped-ion experiment to search for the ground state, and describe that process within a non-equilibrium mean-field approximation. (v) Finally, we provide a set of experimental parameters for the implementation of our model in state-of-the-art trapped-ion setups.

\section{Cooperative Jahn-Teller/ Rabi lattice model in trapped-ion chains}

In this work we deal with a system of spins coupled to a common vibrational bath. Trapped-ion chains will be our reference physical system, although our model can be used to describe a variety of physical setups consisting of quantum emitters coupled to bosonic modes in cavity or circuit QED \cite{Houck12natphys,Schiro2012PRL}.

We consider a chain of $N$ trapped ions with two internal levels, $|0\rangle_j$ , $|1 \rangle_j$, where $j$ is the ion index. 
Internal states are coupled to the radial vibrations of the chain, which
are described by a model of hopping phonons as introduced in many previous works 
(see for example \cite{Porras04prl2}),
\begin{equation}
H_{\rm ph} \left(\{\omega_j \}, \{ t_{j,l} \} \right) =
\sum_j \omega_j a^\dagger_j a_j + \frac{1}{2}
\sum_{j,l} t_{j,l} a^{\dagger}_j a_l,
\label{H.ph}
\end{equation}
where $a_j$ and $\omega_j$ are the annihilation operator of phonons and the local trapping frequencies, respectively. 
Phonons' hopping amplitudes are given by 
\begin{equation}
t_{j,l} = \frac{e^2}{m \omega_x |z_j^{(0)} - z_l^{(0)}|^3}, 
\label{t.def}
\end{equation}
with $e$ the electron charge, $m$ the mass of the ions, $\omega_x$ the radial trapping frequency,  and  $z_j^{(0)}$ the equilibrium position of the $j$ ion. 
Physically, $t_{j,l}$ corresponds to the coupling between dipoles induced by the displacement of the ions relative to their equilibrium positions. 
The vibrational Hamiltonian $H_{\rm ph}$ can be diagonalized in a basis of collective modes $a_j = \sum_{n} M_{j,n} a_n$, so 
\begin{equation}
H_{\rm ph}(\{\omega_n\}) = \sum_n \omega_n a^\dagger_n a_n,
\end{equation}
where ${M}_{j,n}$ and $\omega_n$  are the wave function and frequency of the vibrational mode with index $n$.

A microwave or Raman transition induces a transversal (quantum) field, leading to an effective spin Hamiltonian of the form
\begin{equation}
H_{\rm s} = \frac{\Omega_x}{2} \sum_j \sigma^x_j .
\end{equation}

Finally, we consider a couple of lasers inducing a spin-dependent force,
\begin{equation}
H_{\rm sph}(t) = g \sum_j
\sigma^z_j \left( a_j e^{-i (\Delta k\, z_j^{(0)} + t \delta_j  ) }+ a_j^\dagger e^{i (\Delta k\, z^{(0)}_j +  t \delta_{j} ) } \right),
\end{equation}
where $g$ is the effective spin-phonon coupling. $\Delta k$ is the projection of the effective Raman wave vector induced by the lasers on the chain axis, and $\delta_j= \omega_j - \omega_{\rm L}$ is the (detuning) frequency of the running-wave created by the lasers \cite{Schneider12rpp}. 
We get rid of the time dependence in the spin-phonon coupling by transforming the total Hamiltonian to a rotating frame, such that
\begin{equation}
  \begin{aligned}
	H_{\rm sph}(t) &\to H_{\rm sph}(0),\\
	H_{\rm ph}(\{\omega_j \}, \{t_{j,l}\}) &\to H_{\rm ph} (\{\delta_j \}, \{t_{j,l}\}),
  \end{aligned}
  \label{rotating.frame}
\end{equation}
with effective shifted trapping frequencies $\delta_j$. Our final model has the form of a generalized cooperative Jahn-Teller (or Rabi lattice) Hamiltonian, with dressed spin-phonon couplings,
\begin{equation}
H_{\Delta k}^{\rm JT} = H_{\rm ph} (\{\delta_j \}, \{t_{j,l}\}) + H_{\rm s} + H_{\rm sph}(0).
\label{H.Deltak}
\end{equation}

We focus on homogeneous chains (ions equally spaced by a distance $d_0$), a situation that gives an approximate description of a linear Coulomb crystal, or describes linear arrays of ion microtraps.
The phonon tunneling is expressed in the homogeneous case like \cite{Porras04prl2,Deng08pra},
\begin{equation}
t_{j,l} = \frac{t_{\rm C}}{|j-l|^3}, \ \  t_{\rm C} = \frac{e^2}{m \omega_x d_0^3},
\end{equation} 
and local trapping frequencies $\omega_j$ are constant and approximately equal to $\omega_x$. In the rotating frame defined by the transformation (\ref{rotating.frame}), the final Hamiltonian in the homogeneous limit is given by
\begin{eqnarray}
H_{\Delta k}^{\rm JT} &=& \sum_n \delta_n a^\dagger_n a_n + \frac{\Omega_x}{2} \sum_j \sigma^x_j 
\nonumber \\
&+&  g \sum_{j,n}  \sigma^z_j
\left( M_{j,n} a_n e^{-i \Delta k\,d_0 j} + 
M_{j,n}^* a^\dagger_n e^{i \Delta k\,d_0 j} \right),
\label{JT.h}
\end{eqnarray} 
where $\delta_n = \omega_n - \omega_{\rm L}$ are the collective vibrational frequencies in the rotating frame. A more detailed analysis of the implementation is deferred to Appendix \ref{App.realization}. Hamiltonian (\ref{JT.h}) can be understood as a {\it parent Hamiltonian} for a variety of models describing trapped-ion quantum matter, like effective quantum Ising models.

\section{Undressed ($\Delta k = 0$) Ising Jahn-Teller interactions}

We consider first the case $\Delta k = 0$, corresponding to bare Ising Jahn-Teller interactions. This model has been also introduced in the context of circuit QED \cite{Koch13njp} (Rabi lattice model).

\subsection{Classical limit $\Omega_x = 0$}

We set $\Omega_x = 0$ and study the classical phases of $H_{\Delta k = 0}^{\rm JT}$.
In this case, the Ising Jahn-Teller Hamiltonian can be mapped to an effective Ising model by means of a polaron transformation \cite{Porras04prl1}, 
\begin{equation}
U=e^{-S},\,S = \sum_{j,n} \frac{g}{\delta_n} \sigma^z_j(M_{j,n}a^{\dagger}_n-{\rm H.c.}),
\end{equation}
which decouples bosonic and spin degrees of freedom, so that
\begin{equation}
H_{\Delta k = 0}^{\rm JT}(\Omega_x = 0) \to H_{\rm ph}(\{\delta_n\}) +
 \sum_{j,l} J_{j,l}^{\Delta k = 0} \sigma^z_j \sigma^z_l.
 \label{decoupled.hamiltonian}
\end{equation}

Given any spin eigenstate of this Hamiltonian --let us call it $|\Psi_{\rm spin}\rangle$--, it maps onto the phonons' ground state upon undoing the spin-dependent displacement $U$, so that 
\begin{equation}
|\Psi_{\rm GS}\rangle = |\alpha\rangle |\Psi_{\rm spin}\rangle,\, |\alpha\rangle = e^{\sum_{j,n} \frac{g}{\delta_n} \langle \sigma^z_j\rangle(M_{j,n}a^{\dagger}_n-{\rm H.c.})}|0\rangle.
\label{separable.solution}
\end{equation}

\begin{figure}[h!]
        \begin{subfigure}[ ]
        {\includegraphics[width=2in]{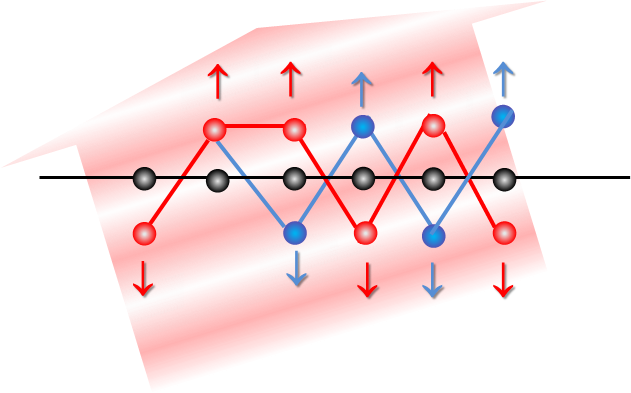} \label{scheme}}
        \end{subfigure}
        \centering
        ~
        \begin{subfigure}[ ]
        {\includegraphics[width=3in]{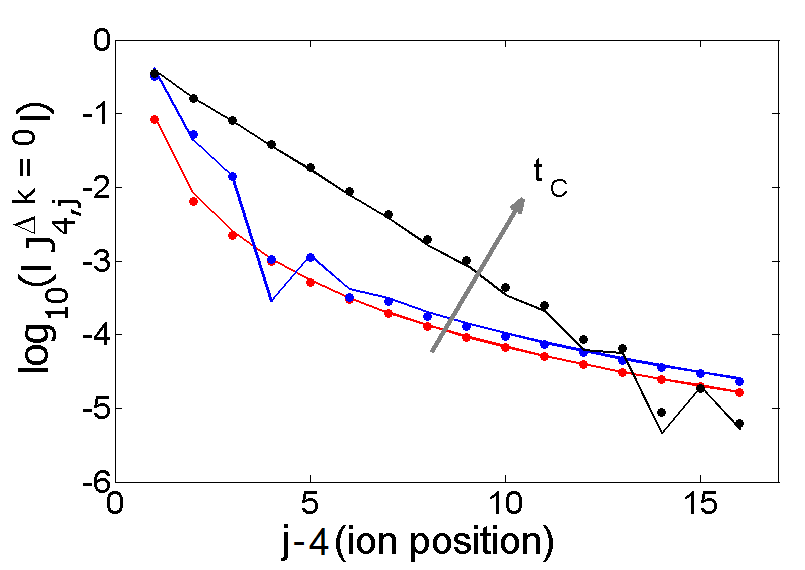} \label{coupling}}
        \end{subfigure}
        \caption{(color online)  
        (a) Configurations induced by dressed spin-phonon couplings: AF (blue) and Hopfield orders (red).
        (b) Effective spin coupling as a function of ion-ion separation in a $N =20$ ion chain, between ion $j = 4$ and the rest of the chain. Energy units such that $\delta_{N/2} = g = 1$, $t_{\rm C} = 0.1, 1, 5$. Circles: exact calculation. Continuous line: estimate from Eq. (\ref{J.scaling}).}
\end{figure}

In this expression, $|0\rangle$ denotes the phonons' vacuum, and $\langle \sigma^z\rangle$ is evaluated upon $|\Psi_{\rm spin}\rangle$. This result expresses that the motional (structural) configuration is inextricably linked to the magnetic order (cf. Fig. \ref{scheme}). This fact has been exploited, for instance, in the quantum simulation of the Jahn-Teller effect with trapped ions \cite{Porras12prl}. 

We now focus exclusively on the Ising term in (\ref{decoupled.hamiltonian}), as it wholly defines the ground state of the chain. The Ising interaction is given by
\begin{equation}
J^{\Delta k = 0}_{j,l} = - 
 \sum_{n}{M}_{j,n} \frac{g^2}{\delta_n}  {M}_{l,n}^*,
\end{equation}
which is determined by the properties of the vibrational states in the rotating frame, $H_{\rm ph}(\{\delta_n\}) = \sum_n \delta_n a^\dagger_n a_n$. We derive an approximate expression in the long chain limit $N \gg 1$ (see Appendix \ref{App.decay}),
\begin{equation}
J^{\Delta k = 0}_{j,l} \simeq - (-1)^{j-l} J_{\rm exp} e^{- \left| j - l \right|/\xi}
+ \frac{J_{\rm dip}}{\left| j - l \right|^3}_{j \neq l},
\label{J.scaling}
\end{equation}
with the constants
\begin{equation}
\left\{
\begin{array}{ll}
\xi = \sqrt{\log{2}/2} \sqrt{t_{\rm C} / \delta_{N/2}},\\
J_{\rm exp} = \frac{\xi g^2}{t_{\rm C} \log{2}},\\
J_{\rm dip} = \frac{g^2 t_{\rm C}}{2(\delta_{N/2} + 7\zeta(3)t_{\rm C}/4)^2}.
\end{array}
\right.
\end{equation}

In the previous expressions, $\delta_{N/2}$ denotes the (detuning) frequency of the laser force from the lowest energy, zigzag vibrational mode, $M_{j,n}=1/\sqrt{N}(-1)^j$. Also, we note that these formulas hold only for $\delta_{N/2}>0$.

The analytical expression in Eq. (\ref{J.scaling}) is very accurate even for moderate $N$ (see Fig. \ref{coupling}). From it, we can read that the effective Ising interaction interpolates between two limits:

(i) Short range limit ($\delta_{N/2} \gg t_{\rm C}$, $\xi \ll 1$).- The effective Ising interaction shows a dipolar decay, $J^{\Delta k=0}_{j,l} \simeq J_{\rm dip} /|j - l|^3$. Effectively, this coupling is first-neighbors. Since $J_{\rm dip}>0$, the classical ground state is antiferromagnetic.

(ii) Long range limit ($\delta_{N/2} \ll t_{\rm C}$, $\xi \gg 1$).- The exponential decay is dominant, and for regions of length $L$, with $L\ll\xi$, it becomes independent of $j,l$. The classical ground state remains antiferromagnetic, because of the alternation of signs in the first term in Eq. (\ref{J.scaling}).

Usually, the onset of long range (effective) spin-spin interactions is associated with making the spin-dependent force nearly resonant with one particular, motional (collective) mode --such as the center-of-mass mode, $M_{j,n}, n=0$--, which features distant spatial correlations \cite{Islam13sci}. These are imprinted in $J_{j,l}^{\Delta k =0}$ through the wave function $M_{j,n}$. This approach may be problematic for $N\gg 1$, since then different modes get closer between each other and are more difficult to resolve individually. However, from Eq. (\ref{J.scaling}), we see that in the event of many ions, the range of $J^{\Delta k = 0}_{j,l}$ relies on the relative values of $\delta_{N/2}$ and $t_{\rm C}$, and not necessarily on some frequency matching condition between the laser force and the motional modes. Furthermore, these parameters are independently tunable. Specifically, $\delta_{N/2}$ depends on the frequency of the force, while $t_{\rm C}$ is fixed by means of the trap frequency $\omega_x$.

\subsection{Quantum regime, $\Omega_x \neq 0$}

The quantum ground state of
\begin{equation}
H^{\rm JT}_{\Delta k = 0} = \sum_n \delta_n a^\dagger_n a_n + \frac{\Omega_x}{2} \sum_j \sigma^x_j 
+g \sum_{j,n}  \sigma^z_j
\left( M_{j,n} a_n+ {\rm H.c.}\right),
\label{quantum.hamiltonian}
\end{equation} 
is determined by the competition between the spin-phonon interaction through the $\sigma^z$-component, and the transversal field, $\Omega_x$. In the regime of strong spin-phonon coupling ($g \simeq \delta_{N/2}$), the system is no longer described by an effective Ising model, since the polaron transformation does not commute with the quantum ($\sigma^x$) term \cite{Porras04prl1}. However, the Hamiltonian (\ref{quantum.hamiltonian}) is invariant under the parity transformation 
\begin{equation}
\sigma_j^z \to -\sigma_j^z, a_j \to -a_j,
\end{equation}
which is analogous to the $\mathbb{Z}_2$ symmetry of the Ising model. Thus, symmetry arguments suggest that the physics of $H^{\rm JT}_{\Delta k = 0}$ should be similar to that of the spin Ising Hamiltonian. In particular, we expect a critical value of $\Omega_{x,\rm c}$ separating a ferromagnetic from a paramagnetic phase.

\begin{figure}[h!]
        \includegraphics[width=3.3in]{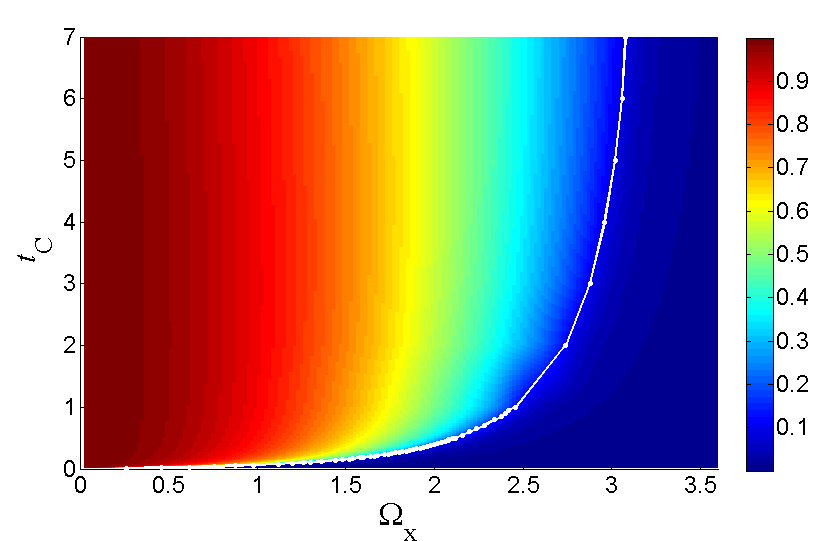} 
        \caption{(color online) DMRG phase map, for $N=40$ sites, in units such that $\delta_{N/2}=g=1$, and $\Delta k = 0$. We depict the expectation value of the OAF (\ref{OAF.parameter}). For the DMRG algorithm we used the parameters $D=22$, number of states kept, and maximum number of phonons per site $n=8$.}
        \label{Fig.Dk0}
\end{figure}

To study this question we have carried out calculations with the DMRG method. In our simulations, we neglect the dipolar character of the vibrational couplings and keep nearest-neighbor interactions only, $t_{j,l} = t_{\rm C} \delta_{j,l+1}$. The phase diagram of the model is presented in Fig. \ref{Fig.Dk0} for a wide range of transversal fields $\Omega_x$ and Coulomb couplings $t_{\rm C}$. Because of the parity symmetry, the $z$-magnetization is zero, so we have to choose a different order parameter. We define the antiferromagnetic order parameter (OAF) as
\begin{equation}
\text{OAF}:=\sum_{j,l} (-1)^{|j-l|}\frac{\langle \sigma_j^z \sigma^z_l \rangle}{N(N-1)}.
\label{OAF.parameter}
\end{equation}
This observable captures the onset of the ordered phase for $\Omega_x < \Omega_{x,\rm c}$, whereas is zero in the paramagnetic phase.

\begin{figure}[h!]
\includegraphics[width=3.3in]{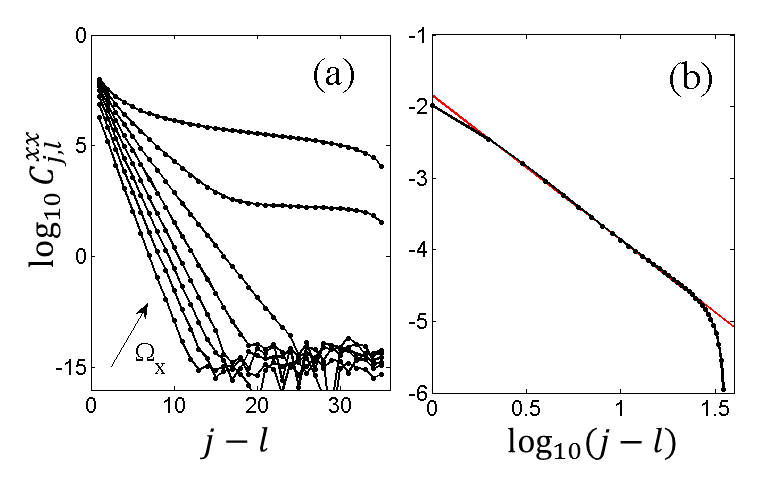} 
        \centering
        \caption{(color online) DRMG correlations $C^{xx}_{j,l}$ for $t_{\rm C} = 0.2$. Same parameters as in Fig. (\ref{Fig.Dk0}). (a) Values $\Omega_x =$ 2 $\times$ 0.1, 0.2, 0.3, 0.4, 0.5, 0.6, 0.7, 0.82. (b) $\log-\log$ plot for $\Omega_x = \Omega_{x, \rm c} =$ 1.64. We fit the result to a line, resulting in a slope $\nu = $ 2.02.
        }
        \label{corr2}
\end{figure}

The calculations show that there is a second order phase transition for a critical value, $\Omega_{x, {\rm c}}$ of the transversal field. In particular we have studied the behavior of correlation functions,
\begin{equation}
C^{xx}_{j,l} = \langle \sigma^x_j \sigma^x_l \rangle - 
\langle \sigma^x_j \rangle \langle \sigma^x_l \rangle .
\end{equation}
We see that the $C^{xx}_{j,l}$ decay exponentially for values 
$\Omega_{x} \neq \Omega_{x, {\rm c}}$ and show a power-law decay,
$C^{xx}_{j,l} \sim  |j - l|^{-\nu}$ for $\Omega_{x} = \Omega_{x, {\rm c}}$, with $\nu \simeq 2$, consistent with a phase transition within the quantum Ising universality class (cf. Fig. \ref{corr2}).
By calculating the points in the phase diagram for which the correlation functions have the longest range we are able to identify the critical line shown in Fig. \ref{Fig.Dk0}. 

\begin{figure}[h!]
        \includegraphics[width=3.3in]{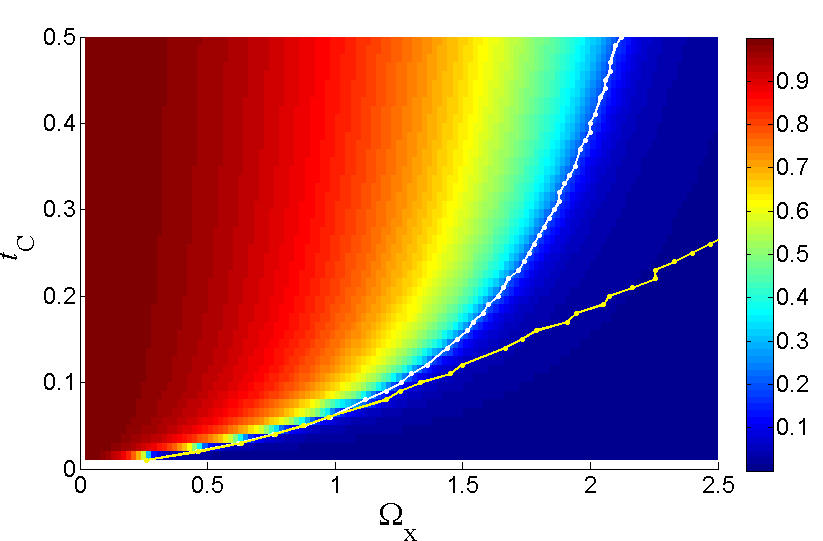} 
        \caption{(color online) OAF values for $t_{\rm C}\to 0$. We show the DMRG (upper line) vs. the short range predictions (\ref{short.range.prediction}) (bottom line). There is agreement in the short range limit $t_{\rm C}\ll \delta_{N/2} = 1$. The rest of the parameters are the same as in Fig. (\ref{Fig.Dk0}).}
        \label{Fig.Dk0.2}
\end{figure}

\begin{figure}[h!]
\includegraphics[width=3.3in]{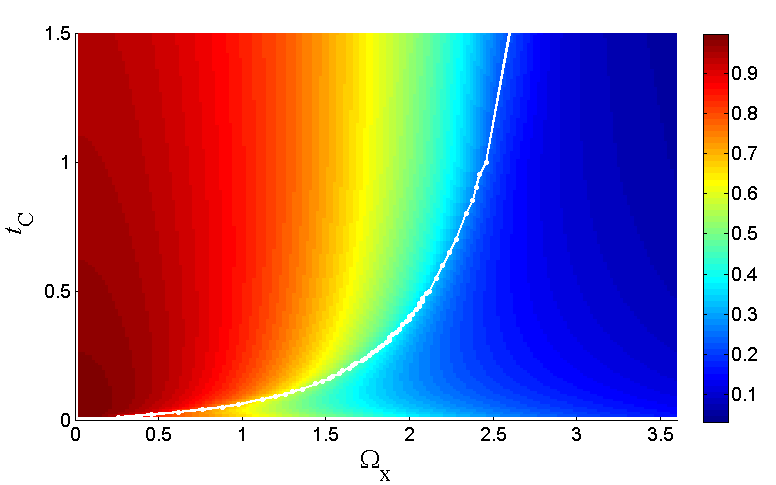}
        \centering
        \caption{(color online) DMRG mean phonon number. Same parameters as in Fig.(\ref{Fig.Dk0}).}
		\label{nph.Dk0}
\end{figure}

We can estimate the scaling of the critical line in the limit of $\delta_{N/2} \gg t_{\rm C}$. In this case, our problem can be described by an effective Ising model with critical field
\begin{equation}
\Omega_{x, {\rm c}} = \frac{g^2}{\delta_{N/2}} e^{-2 \bar{n}}.
\label{short.range.prediction}
\end{equation}
This expression correctly predicts the behavior of the critical line for $t_{\rm C} \to 0$, as can be seen in Fig. \ref{Fig.Dk0.2}. It depends on the numerical values of the mean phonon number $\bar{n} = 1/N \sum_{j=1}^{N} \langle a^{\dagger}_na_n \rangle$ (cf. Fig. \ref{nph.Dk0}).

Generally, for $\Omega_x \neq 0$, phonons may be entangled with spins in a more convoluted way than the separable solution (\ref{separable.solution}). Nevertheless, we know that this state is exact in the limits of $\Omega_x=0$ and $\Omega_x\to\infty$. Also, it has been shown that this description applies, at least qualitatively, to the regime of long range interactions $\xi\gg 1$ \cite{Nevado13epjst}. This last claim is supported by the increasing degree of correlation for larger $t_{\rm C}$ between the mean phonon number $\bar{n}$ (Fig. \ref{nph.Dk0}) and the OAF parameter (Fig. \ref{Fig.Dk0}). The change of $\bar{n}$ at $\Omega_x = \Omega_{x,\rm c}$ signals the quantum magnetic/structural equivalent of the zigzag structural phase transition \cite{Fishman08prb,Montangero14prb}.

\section{Dressed ($\Delta k \neq 0$) Ising Jahn-Teller Interactions}

Now we discuss the effect of non-zero optical phases of the laser force. In this case, $\Delta k$ dresses the (effective) spin-spin interactions, so as 
\begin{equation}
J^{\Delta k}_{j,l} = \cos[\Delta k\,d_0(j-l)]\, J^{\Delta k=0}_{j,l}.
\label{J.dressed}
\end{equation}

The periodically alternating strengths of the spin-spin couplings stem from the laser force inhomogeneous phases, which imprint an extra spatial modulation on the collective motion of the chain. We shall see that this feature, together with the tunable range of $J^{\Delta k}_{j,l}$, gives rise to frustration in the system.

\subsection{Classical limit $\Omega_x = 0$}

We study the case $H^{\rm JT}_{\Delta k}(\Omega_x=0)$. As in the previously considered classical limit, the motional ground state consists of ions oscillating in the structural configuration dictated by the local spin projections along the z axis. The spin ground states are determined by the range of the effective, dressed interactions $J_{j,l}^{\Delta k}$. Accordingly, we distinguish between:

(i) Short range limit ($\delta_{N/2} \gg t_{\rm C}$, $\xi \ll 1$).- Here the ground state is either antiferromagnetic (AF) or ferromagnetic (F), depending on the sign of 
the nearest-neighbor couplings: 
$\langle \sigma^z_j \rangle_{\rm F} = \langle \sigma^z_{j+1} \rangle_{\rm F}$ if 
$J^{\Delta k}_{j,j+1} < 0$, and
$\langle \sigma^z_j \rangle_{\rm AF} = - \langle \sigma^z_{j+1} \rangle_{\rm AF}$ if 
$J^{\Delta k}_{j,j+1} > 0$. 

(ii) Long range limit ($ \delta_{N/2} \ll t_{\rm C} $, $\xi \gg 1$).- Now we can approximate
\begin{equation}
J_{j,l}^{\Delta k} = 
- J_{\rm exp} \sum_{m = c,s} \chi^{[m]}_j  \chi^{[m]}_l,
\label{J.hopfield}
\end{equation}
with
$\chi^{[c]}_j = (-1)^j \cos(\Delta k\,d_0 j)$,
$\chi^{[s]}_j = (-1)^j \sin(\Delta k\,d_0 j)$.
These two different configurations of the couplings can be independently satisfied, because they are (quasi-) orthogonal. Thus, both solutions $\langle \sigma^z_j \rangle_{\rm Hopf} = {\rm sign}(\chi^{[c]}_j)$, and  $\langle \sigma^z_j \rangle_{\rm Hopf} = {\rm sign}(\chi^{[s]}_j)$ are ground states of the problem, and they are degenerate when $N\to \infty$. We have obtained an effective model reminiscent of Hopfield's description of neural networks \cite{Hopfield82PNAS,Pons07prl}. The magnetic order is set by the laser force inhomogeneous phase, and consists of the two possible arrangements $\{{\rm sign}(\chi^{[c]}_j), {\rm sign}(\chi^{[s]}_j)\}$.

This situation changes drastically in between the short and long range limits. In this regime, that lies around $t_{\rm C}=\bar{t}_{\rm C}\simeq  0.5$ for the parameters we are using, the exponential and dipolar contributions to $J_{j,l}^{\Delta k}$ have similar weights, and the minimum energy configuration must be an interplay between the AF/F and Hopfield ground states. Since these two sets of orders cannot be simultaneously satisfied, there are many spins for which pointing either upwards or downwards has no contribution to the ground state energy. As a consequence, a huge number of different states have energies very close to the global energy minimum. This large amount of (quasi-) degeneracy is the defining characteristic of frustration.

\subsection{Quantum regime, $\Omega_x \neq 0$}

In this case, we are going to focus exclusively on the spin degrees of freedom. The onset of magnetic frustration around $t_{\rm C}=\bar{t}_{\rm C}$, together with a non-zero value of $\Omega_x$, has dramatic consequences for the ground state. The large amount of (quasi-) degenerate, classical ground states, gives rise to a huge set of coherent superpositions in the quantum regime, leading to the enhancement of quantum fluctuations, and to the increase of the quantum correlations. These two elements are a landmark of the spin-liquid phases in frustrated magnets \cite{balents10nat,Ramirez94}. 

\begin{figure}[h!]
\includegraphics[width=1\linewidth]{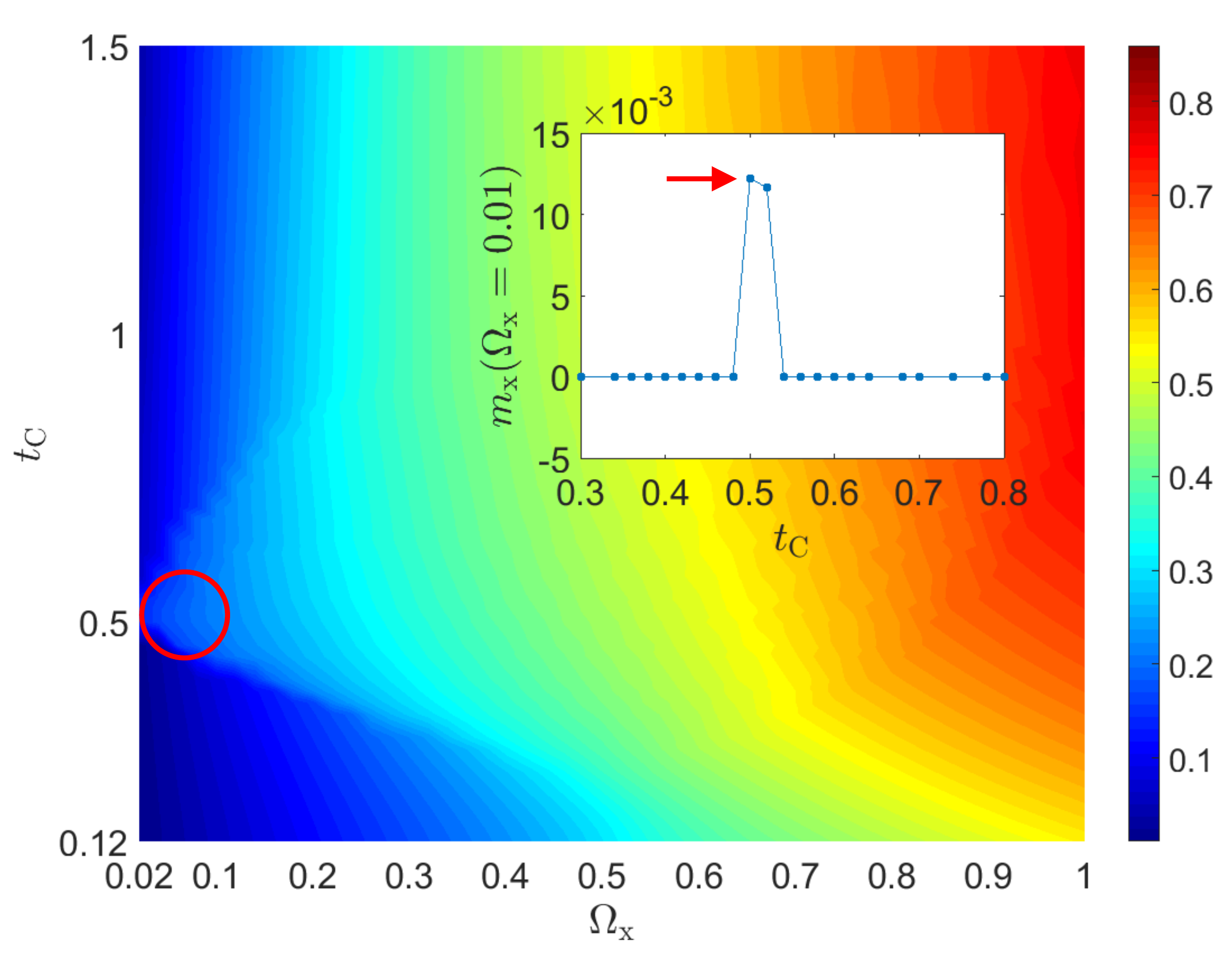}
\caption{(color online) Quantum phase diagram of an ion chain with $N=40$ ions. Energy units $\delta_{N/2}=g=1$. We show the DMRG calculations for $\Delta k = 2 \pi /(3d_0)$. Rest of parameters as in Fig. (\ref{Fig.Dk0}). The red circle pinpoints the highly frustrated regime. The inset shows the sudden enhancement of quantum fluctuations between the ordered and frustrated phases.}
\label{quantum.phase.map}
\end{figure}

\begin{figure*}[t!]
\includegraphics[width=1\linewidth]{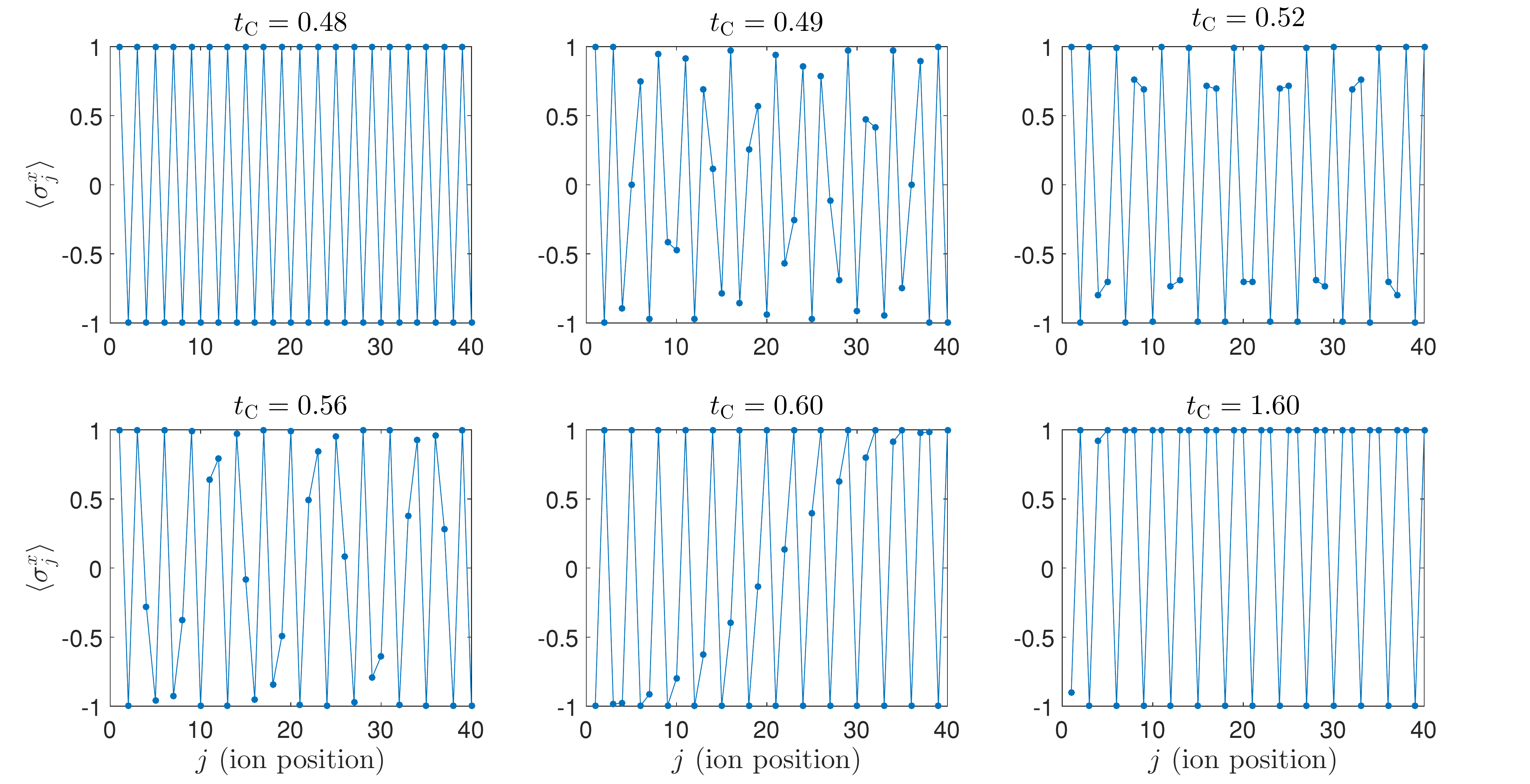} 
\caption{(color online) Different ground states for $\Delta k = 2 \pi/(3 d_0),\Omega_x = 0.01$. The rest of the parameters are the same as in Fig. (\ref{Fig.Dk0}).  For $t_{\rm C} = 0.48$ the system is in the AF phase. It suddenly jumps into the frustrated regime for $t_{\rm C} = 0.49$, and features different exotic configurations in the highly frustrated regime, around the value $\bar{t}_{\rm C} \simeq 0.5$. Finally, for $t_{\rm C} = 1.6$ the system is well inside the non-frustrated Hopfield phase.}
\label{frustrated.orders}
\end{figure*}

Quantum fluctuations in the frustrated regime are unveiled by the increase of the mean $x$-magnetization, $m_{\rm x} = (1/N) \sum_j \langle \sigma^x_j \rangle$. As many spins are subject to frustrated interactions, they must align appreciably with $\sigma^x$ for infinitesimal values of $\Omega_x$. Outside the frustrated regime, we recover either the AF/F or the Hopfield orders, which are more resilient against perturbations. Therefore, we see a sudden change in $m_{\rm x}$ depending on whether the system is in the frustrated region or not (cf. Fig. \ref{quantum.phase.map}). Indeed, we have calculated some spin orders for different values of $t_{\rm C}$, that illustrate the transition between the ordered and the frustrated phases (Fig. \ref{frustrated.orders}).

\begin{figure}[h!]
\includegraphics[width=0.8\linewidth]{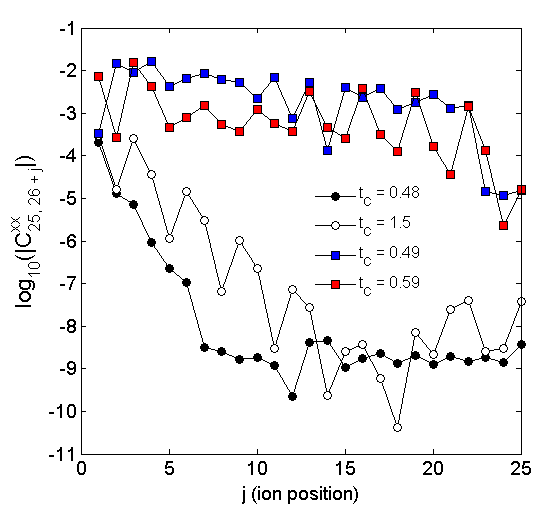} 
\caption{(color online) Results for the ground state of a chain with $N = 50$ ions calculated with a DMRG algorithm (number of states kept $D$ = 18, maximum number of phonons $n = 9$), in units of $g = 1$, $\delta_{N/2} = 1$, with $\Omega_x = 0.01$, and $\Delta k = 5 \pi / (3 d_0)$. Correlations functions 
$C^{\rm xx}_{j,l} = \langle \sigma^x_j \sigma^x_l \rangle - \langle \sigma^x_j \rangle \langle \sigma^x_l \rangle$ for values of $t_{\rm C}$ in the AF/F, Hopfield and frustrated phases.}
\label{Corr_frustration}
\end{figure}

For the calculations of the spatial quantum correlations in this case, we have performed an independent simulation. This is depicted in Fig. \ref{Corr_frustration}. We see that $C^{xx}_{j,l}$ exhibits the typical exponential decay of gapped phases \cite{Sachdev.book}, for the short and long range coupling regimes. The correlations in the AC/F phase ($t_{\rm C}=0.48$), exhibit no structure apart from the exponential decay. On the contrary, in the Hopfield phase ($t_{\rm C}=1.5$), the periodicity imprinted by the force inhomogeneous phase $\Delta k = 5\pi/(3d_0)$ shows up clearly on the spatial correlations. However, the most interesting effect is the enhancement of correlations in the highly frustrated regime. The abrupt change in correlation range between $t_{\rm C}=0.48$ (no frustration) and the value $t_{\rm C} = 0.49$ (frustrated region) shows that the variation on the physical properties of the ground state between these two regimes happens very sharply. This provides us with another accurate resource to pinpoint the frustrated regime, along with the enhancement of quantum fluctuations.

\section{Quantum annealing}
\label{quantum.annealing}

Our model can be implemented with trapped ions, and used to assess the efficiency of quantum annealing \cite{Finnila94cpl,nishimori98pre} in finding a global classical ground state in the presence of frustrated interactions \cite{Nishimori15jpa}.
Here we resort to a mean-field ansatz to predict the quasi-adiabatic evolution of the system. 
Our method is very similar to the one recently used in the theoretical description of quantum annealers fabricated with superconducting qubits \cite{Smolin14arX}. 
As we shall see this approach is not fully conclusive about the performance of the quantum protocol. 
Still, our calculation is interesting since trapped-ion experiments can shed light on the relevance of quantum correlations in quantum annealing, by means of a comparison between  the output of analogical quantum simulations and our classical mean-field results.

We devise a protocol described by the Hamiltonian 
\begin{equation}
H_{\rm ad}(t) = H_{\rm ph} (\{\delta_n\}) + H_{\rm s} (\Omega_x(t)) + \frac{1}{2}\Omega_{ z}(t) \sum_j \sigma^z_j + H_{\rm sph} (g(t)),
\label{adiabatic.protocol}
\end{equation}
where the parameters evolve as
\begin{equation}
\begin{aligned}
\Omega_x(t)& = \Omega_x(0) e^{- t /\tau_{\rm ev}}, \\
\Omega_{ z}(t) &= \Omega_{ z}(0) e^{- t /\tau'_{\rm ev}}, \\
g(t)^2 &= g(0)^2 (1 - e^{-t/\tau_{\rm ev}}).
\end{aligned}
\label{schedules}
\end{equation}

We note that $H_{\rm ad}(t)$ consists of the original $H_{\Delta k}^{\rm JT}(t)$ --that now changes in time according to the annealing schedules $g(t),\Omega_x(t)$--, and a longitudinal field of magnitude $\Omega_z(t)$. In a real experiment, one can dispose of this latter term at no expense. However, as we explain below, this symmetry-breaking field is necessary for the mean-field treatment of the dynamics. Also, we consider two different typical time scales $\tau_{\rm ev},\,\tau'_{\rm ev}$ in the evolution. To optimally hit the classical ground state, these times must fulfill $\tau'_{\rm ev} \ll \tau_{\rm ev}$. This is because a non-zero value of $\Omega_z$ modifies the target energy landscape of the problem. Therefore, $\Omega_z$ must be effectively zero at the time the (quantum) field $\Omega_x$ is performing the quantum annealing. We accomplish this requirement by imposing the former condition upon the two different typical times.

The annealing process proceeds as follows. According to (\ref{schedules}), at $t = 0$ there is no spin-phonon coupling.  We initialize the dynamics in the state
\begin{equation}
|\Psi_{\rm GS}(t=0)\rangle =|0\rangle \bigotimes_{j=1}^N \left |\theta_j\right\rangle.
\label{initialization.state}
\end{equation} 
The vector $\left |\theta_j\right\rangle$ is an eigenstate of the fields along $x$ and $z$ directions. Next, the system evolves towards the strongly coupled spin-phonon regime $\Omega_x,\Omega_z \to 0, g\neq 0$, and at some time $t_f\gg \tau_{\rm ev}, \tau'_{\rm ev}$, it reaches a final state, that under ideally adiabatic evolution would correspond to the classical ground state.

To test this picture, we solve the dynamics described by (\ref{adiabatic.protocol}) within a mean-field approximation. Specifically, we derive the Heisenberg equations of motion for the expectation values of $a_n(t), \sigma_j^{\alpha}(t),\alpha =x,y,z$ upon a particular vector $|\Psi_{\rm GS}\rangle$. We assume that this vector is a separable state of bosons and spins, and that these latter are decoupled between neighboring sites. This reduces every expectation value of the product of observables to the product of their mean values.

This approximation must be handled with care, as it can lead to incorrect predictions. For instance, let us assume that $\Omega_z (t)=0,\forall t$. In this case, the initial condition for the dynamics is no longer (\ref{initialization.state}), rather $|\Psi_{\rm GS}'(t=0)\rangle =|0\rangle \bigotimes_j \left |\downarrow_{x,j}\right\rangle$. As a consequence of the separability assumption, it turns out that this state is a stationary solution of the Heisenberg equations. However, this is not correct, as the exact dynamics would transfer population away from this state, owing to the spin-boson coupling. This is the reason for including the term $\Omega_z$ in (\ref{adiabatic.protocol}).

\begin{figure}[b!]
\includegraphics[width=0.5\textwidth]{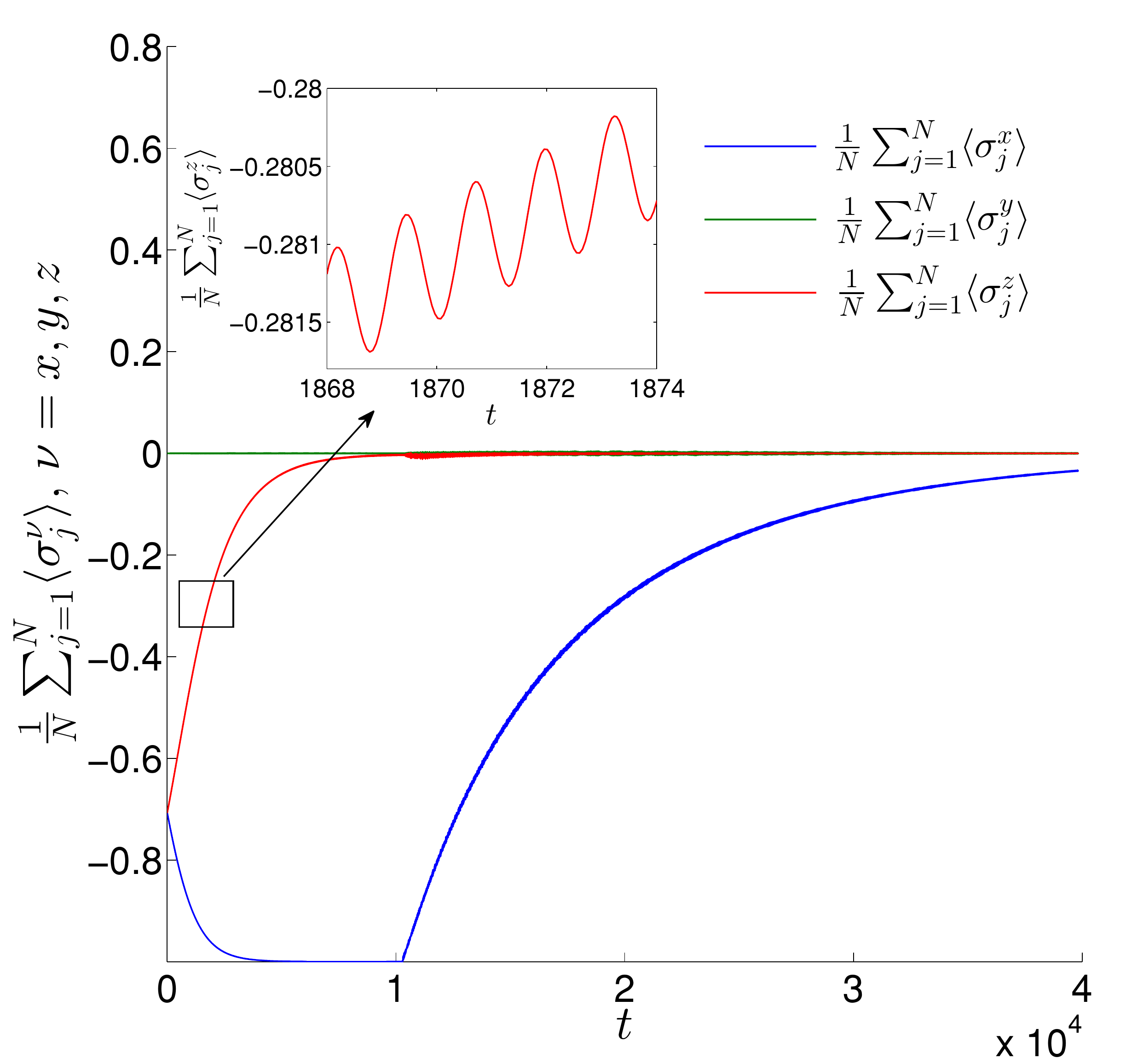}
\caption{(color online) Evolution of the spin expectation values. We set $t_{\rm C}=2$, for units such that $g=\delta_{N/2}=1$, with $\Delta k=0,\Omega_x=\Omega_z=5, \tau_{\rm ev}=8\tau_{\rm ev}'=10^{-4}.$}
\label{ev2}
\end{figure}

Our protocol is intended to work in the strongly coupled spin-phonon regime, where bosonic degrees of freedom cannot be adiabatically eliminated. Nevertheless, it is still possible to encode the whole dynamics exclusively in the spin degrees of freedom, as an artifact of the mean-field approximation. Let us write the equation of motion for the collective phonon operators in the interaction picture with respect to $H_{\rm ph}(\{\delta_n\})$
\begin{equation}
\frac{d}{dt} \langle \tilde{a}_n (t)\rangle = 
-ig\left(\frac{t}{\tau_{\rm ev}}\right) \sum_j \langle \sigma_{j}^z (t) \rangle e^{i\Delta k j} M_{j,n}^* 
e^{i{\delta}_n t},
\end{equation}
This interaction picture amounts to the change of variable $\tilde{a}_n = a_n e^{i {\delta}_n t}$, where the $a_n$ are the phonon operators in the Schr{\"o}dinger picture. Since the protocol starts in the paramagnetic phase, where $\langle 
\tilde{a}_n (0)\rangle=0$, we get, after an integration by parts,
\begin{align}
&\langle \tilde{a}_n (t) \rangle =  
-\frac{g\left(\frac{t}{\tau_{\rm ev}}\right)}{{\delta}_n} \sum_j 
\langle \sigma_{j}^z (t) \rangle 
e^{i\Delta k j} M_{j,n}^* e^{i{\delta}_n t}  \nonumber \\
&+ \int_{0}^t dt' \frac{\frac{d}{dt''}\left[ g\left(\frac{t}{\tau_{\rm ev}}\right) \sum_j \langle \sigma_{j}^z (t'') \rangle e^{i\Delta k j} M_{j,n} \right]_{t''=t'}}{{\delta}_n} e^{i{\delta}_n t'}.
\label{adiabatic}
\end{align}
We can drop the second term in the right hand side of Eq. (\ref{adiabatic}) 
as long as we assume that 
$d_t \langle \sigma_j^x (t) \rangle \ll{\delta}_n$ and  $\tau_{\rm ev}^{-1}, \tau_{\rm ev}'^{-1} \ll 
{\delta}_n$. The latter requirement is fulfilled for long enough evolution times
$\tau_{\rm ev}$, $\tau'_{\rm ev}$, but the first condition must be validated a posteriori from the self-consistency of the results (cf. Fig. \ref{ev2}). 

We plug this last relation into the remaining equations of motion, and finally we arrive to the closed set
\begin{equation}
\begin{aligned}
\frac{d}{dt}\langle \sigma_{j}^x (t)\rangle &= -\Omega_z(0) e^{-\frac{t}{\tau_{\rm ev}'}} \langle \sigma_{j}^y (t)\rangle \\
&+ 2\sum_l (1-e^{-\frac{t}{\tau_{\rm ev}}}) J^{\Delta k}_{j,l} \langle \sigma_{j}^y (t)\rangle\langle \sigma_{l}^z(t)\rangle,
\\
\frac{d}{dt}\langle \sigma_{j}^y (t)  \rangle &= \Omega_z(0) e^{-\frac{t}{\tau_{\rm ev}'}}\langle \sigma_{j}^x (t)\rangle-\Omega_x(0) e^{-\frac{t}{\tau_{\rm ev}}} \langle \sigma_{j}^z (t)\rangle  \\
&-2\sum_l (1-e^{-\frac{t}{\tau_{\rm ev}}}) J^{\Delta k}_{j,l} \langle \sigma_{j}^x (t)\rangle\langle \sigma_{l}^z (t)\rangle,
\\
\frac{d}{dt}\langle \sigma_{j}^z (t)\rangle &= \Omega_x(0) e^{-\frac{t}{\tau_{\rm ev}}} \langle \sigma_{j}^y (t)\rangle.
\end{aligned}
\end{equation}

\begin{figure}[h]
\centering
\vspace{0.1in}
\includegraphics[width=.5\textwidth]{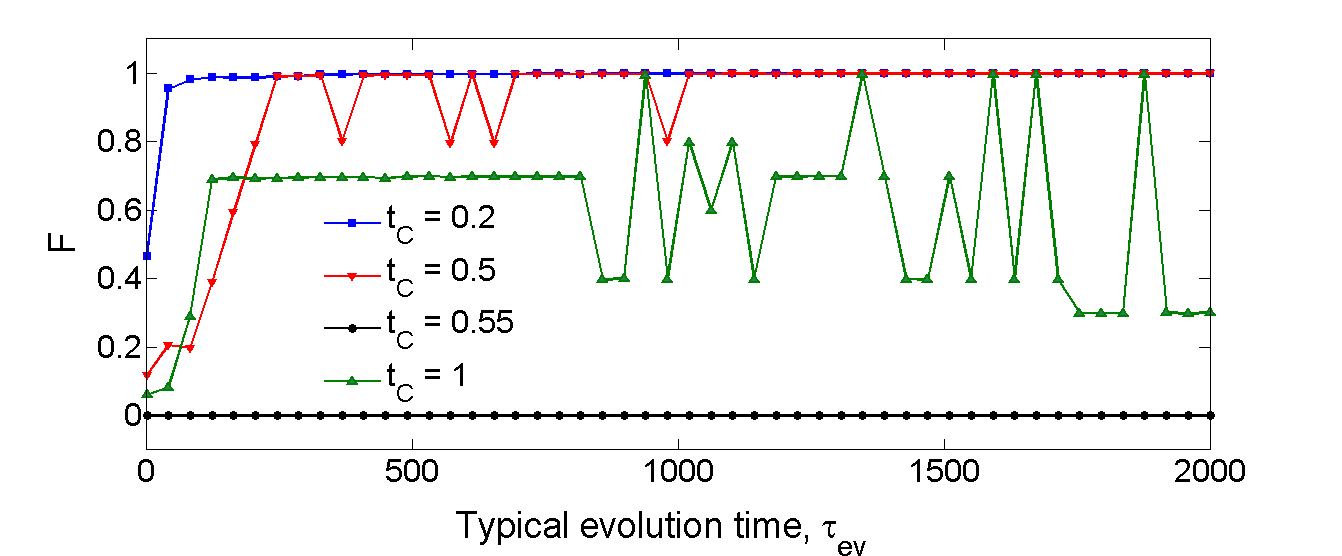}
\caption{(color online) Outcome of the quantum annealing protocol. $N=20$, in energy units such that $\delta_{N/2}=g=1$, and values of $t_{\rm C}$ as indicated in the figure. $\Delta k = 2\pi/(3 d_0)$, and $\Omega_x(0) = 5 $, $\Omega_z(0) = 10^{-1}~\Omega_x(0)$, and $\tau_{\rm ev} = 10~\tau_{\rm ev}'$.}
\label{fig.adiabatic}
\end{figure}

We now integrate these equations, and present the results in Fig. \ref{fig.adiabatic}. We calculate $\langle \sigma^z_{j}(t)\rangle$, and at a time $t = t_{{\rm f}} \gg \tau_{\rm ev},\tau_{\rm ev}'$, we quantify the fidelity between the exact ground state in the classical limit, $\langle \sigma_j^z \rangle_{\rm ex}$, and the quantum annealing result, $\langle \sigma_j^z \rangle_{\rm QA}$, by the overlap, 
\begin{equation}
F = \frac{1}{N} | \sum_j \langle \sigma_j^z \rangle_{\rm ex} \langle \sigma_j^z \rangle_{\rm QA} |.
\end{equation}
Finally, this calculation is performed for increasing typical evolution times $\tau_{\rm ev}$, and for different values of $t_{\rm C}$, in both the ordered and frustrated regimes.

We recall that for $\Omega_x = 0$, the whole ground state is completely defined by the spin configuration compatible with $J_{j,l}^{\Delta k}$. In the presence of frustration, finding such configuration requires sampling among a huge number of candidates. Therefore, we had to restrict our simulation to $N=20$ ions, in which case the exact spin ground states can be computed by usual minimization routines.

The mean-field adiabatic evolution of the spin-phonon system is predicted to reach the global ground state far from the highly frustrated regime. Accordingly, for values of the coupling $t_{\rm C} \ll \bar{t}_{\rm C}$, the evolution hits the F phase (upper line in Fig. \ref{fig.adiabatic}). We recall that this is the ground state in the short range interaction regime in the case of $\Delta k = 2\pi/(3d_0)$. Quite remarkably, when $t_{\rm C}=\bar{t}_{\rm C}\simeq 0.5$ (second line from the top), the mean-field dynamics is still able to reproduce the exact solution. The fact that it takes a longer time to converge signals the onset of a more convoluted energy landscape. We note as well that before reaching the value $F=1$ steadily, the protocol explores other orders, pinpointing the existence of quasi-degenerate solutions. On the contrary, deeper in the highly frustrated regime $t_{\rm C} = 0.55$ (bottom line), the mean-field ansatz performs very poorly, and it does not abandon the initial condition for any of the time scales considered. For $t_{\rm C} = 1$ (third line from the top), the simulation is also unable to converge in the time scales considered. These two latter results may imply that in this regime there are features that cannot be captured by the mean-field ansatz. 

We stress that only an exact simulation of the full, quantum dynamics, would assess the validity of quantum annealing as an algorithm to find the lowest energy configuration in this system. However, it has been argued that a mean-field approach, as the one we have studied, suffices to capture the output of a physical realization of our protocol \cite{Smolin14arX}. This amounts to neglect the effect of quantum correlations on the exact dynamics. In the light of the previous results, we cannot claim that the mean-field approach correctly predicts the exact annealing dynamics, nor that it does not. Therefore, this system is an ideal scenario for experiments to test whether a quantum correlated annealer would yield results which cannot be described by our mean-field (separable) ansatz.

Regarding the condition $d_t \langle \sigma_j^x (t) \rangle \ll{\delta}_n$, we present in Fig. \ref{ev2} the evolution of the spin mean values during the annealing process. The inset shows the spin precession as a result of the asymmetry in the typical times of evolution between the field $\Omega_x$ and the symmetry breaking term $\Omega_z$. We estimate $d/dt \langle \sigma_j^z (t) \rangle$ as the product of the amplitude of the oscillation divided by its period, so that $d/dt \langle \sigma_j^z (t) \rangle \sim 10^{-3}\ll \min_n {\delta}_n=\delta_{N/2}=1$. Therefore, we see that dropping the integral term in Eq. (\ref{adiabatic}) is justified in this case.

\section{Trapped ions experimental parameters}
Our model $H_{\rm JT}^{\Delta k}$ can be naturally implemented with state-of-the-art trapped-ion setups. Initialization of the system would include cooling the chain close to the vibrational ground state.
A typical average distance between ions is $d_0 =$ 10 $\mu$m. Considering $^9$Be$^+$ ions and a radial trapping frequency $\omega_x$ = 5 (2$\pi$) MHz, we get a Coulomb coupling $t_{\rm C} =$ 98 kHz. The latter is the most important experimental energy scale, limiting the overall speed of an experiment. Axial trapping frequencies for $N =$ 20 and 50 ions with those parameters are $\omega_z =$ 192 and 94 (2$\pi$) kHz, respectively. To induce the spin-dependent force, Eq. (\ref{H.Deltak}), a running wave detuned from the first red sideband transition by $\omega_{\rm x} - \delta_{N/2}$ may be applied. The detuning can be chosen so as to yield a range of values for $t_{\rm C}/\delta_{N/2}$ that covers the phase diagram in Fig. \ref{Fig.Dk0}. Spin-phonon couplings, $g$, can be in the range of 100 kHz, thus comparable to $t_{\rm C}$, \cite{Leibfried03rmp}. Illuminating the ions with an optical force with effective wavelength $\lambda \simeq$ 320 nm, would require a small misalignment with the direction transversal to the ion chain axis with angle $\theta \simeq 0.6$ degrees to render the value $\Delta k = 2 \pi /(3 d_0)$ used in the examples above. With those values, the evolution time for quantum annealing in Fig. \ref{fig.adiabatic} would be of the order of ms. Radial and axial vibrational normal modes cannot overlap to avoid exciting axial modes with the spin-dependent force, a condition that is fulfilled in the range of parameters considered here. We remark that the strongly coupled spin-phonon regime considered here (with $\delta_{N/2} \simeq g$), allows us to avoid the condition $g \ll \delta_{N/2}$ \cite{Porras04prl1}.

\section{Conclusions and outlook}
We have shown that one of the simplest models that can be experimentally implemented with trapped ions yields a fascinating phenomenology due to the interplay between long range interactions and frustration induced by optical phases. Our model can be extended to other setups, like for example arrays of superconducting qubits coupled to microwave cavities, where a similar dressing of qubit-boson couplings has recently been proposed in \cite{Quijandria13prl}. Our proposal can be used to experimentally explore the efficiency of quantum annealing in a controllable one-dimensional setup, thus shedding light on applications of this method in related setups \cite{Boixo13natphys,Smolin14arX}.

\section*{Acknowledgments}

We acknowledge the EU Marie Curie C.I.G. People Programme (Marie Curie Actions) of the European Union’s Seventh Framework Programme (FP7/2007-2013, REA grant agreement no: PCIG14- GA-2013-630955), and the COST Action MP1001 - IOTA.

\appendix
\section{Dressed trapped-ion spin-phonon chain}
\label{App.realization}

\noindent Our model can be realized either in a linear Paul trap, where the spatial distribution of the chain is inhomogeneous, or in a linear array of microtraps, where the separation between ions can be made homogeneous by fabrication. We consider here in more detail the most common case of a linear crystal in a Paul trap, with $N$ ions of charge $e$ and mass $m$ along the trap axis $z$. 
The ions repulsive interaction along with the effective quadratic potential of the trap --with magnitudes $\omega_{x}, \omega_{z}$ for the transversal and longitudinal components respectively-- render a crystal structure which forms a chain for strong radial confinement $\omega_{x} \gg \omega_{z}$ \cite{Deng08pra}. 

The position of each ion $j$, can be described like
\begin{equation}
	\mathbf{r}_j = \delta x_j ~\mathbf{\hat {x}}_j+(z_{j}^{(0)} +\delta z_{j})~\mathbf{\hat {z}}_j,
	\label{coordinate}
\end{equation}
where $z_{j}^{(0)}$ are the equilibrium positions along the chain and $\delta x_{ j}$ and $\delta z_{ j}$ are the displacement operators in the radial and axial direction, respectively. A Taylor expansion of the Coulomb potential around the equilibrium positions up to second order leads to an harmonic Hamiltonian. In the harmonic approximation, vibrations in different directions are not coupled. The Hamiltonian describing the radial vibrations reads \cite{Deng08pra}:
\begin{eqnarray}
	H_{\rm ph} &=& \frac{1}{2m}\sum_j p_{x,j}^2  + \frac{1}{2}m \omega_{x}^2 \sum_j \delta x_{j}^2\nonumber\\
	&-& \frac{1}{2} \sum_{j>l} \frac{e^2}{|z_{j}^{(0)}-z_{l}^{(0)}|}(\delta x_{j}-\delta x_{l})^2.
\end{eqnarray}
We neglect the axial displacements, since they will be weakly coupled to the spin-dependent forces introduced below. Now we second quantize this expression by means of the identifications $\delta x_{j} = 1/\sqrt{2 m \omega_{x}}(a_{j} + a^{\dagger}_{j})$ and $p_{x, j} = i \sqrt{m \omega_{x}/2}(a^{\dagger}_{j}-a_{j} )$. Furthermore, we dismiss the terms that do not conserve the phonon number such as  $a_{j}^2$ or $(a_{j}^{\dagger})^2$ because they are fast rotating as long as $\omega_{x} \gg t_{j,l}$, which in turn is fulfilled because the harmonic potential energy must be greater than the Coulomb energy. Then we are arrive to the final Hamiltonian for the oscillations
\begin{equation}
	H_{\rm ph} \left(\{\omega_j \}, \{ t_{j,l} \} \right)=  \sum_j \omega_{j} a^{\dagger}_{j} a_{j}+\frac{1}{2}\sum_{j>l} t_{j,l} (a^{\dagger}_{j} a_{l} +{\rm H.c.}),
\end{equation}
with the on-site frequency and Coulomb mediated long range hopping given by
\begin{equation}
	\omega_{j}=\omega_{x} - \frac{1}{2}\sum_{p \neq j} \frac{e^2}{m\omega_{x} |z_{j}^{(0)}-z_{p}^{(0)}|^3},\qquad t_{j,l}=\frac{e^2}{m\omega_{x} |z_{j}^{(0)}-z_{l}^{(0)}|^3}.
\end{equation}

The spin-dependent force can be realized by two lasers inducing an AC Stark shift between two effective levels with $\Delta \mathbf{k} = \mathbf{k}_1-\mathbf{k}_2$ and $\omega_{\rm L}=\omega_1-\omega_2$
\begin{equation}
	H_{\rm sph} (t)= \frac{\Omega_{\rm ac}}{2} \sum_j \sigma_j^z \cos(\Delta \mathbf{k \cdot r}_j-\omega_{\rm L} t)
	\label{spin.phonon}
\end{equation}
where we can approximate the product $\Delta {\bf k} \cdot {\bf r}_j$ in terms of the equilibrium positions and displacements,
\begin{eqnarray}
	\Delta \mathbf{k \cdot r}_j&=& (\Delta k_x,0,\Delta k_z)\cdot (\delta x_{j},0,z_{j}^{(0)} +\delta z_{j})\nonumber\\
	&=&\eta_x(a_{x,j} e^{-i\omega_x t}+a_{x,j}^{\dagger} e^{i\omega_x t}) + \eta_z(a_{z,j} e^{-i\omega_z t}+a_{z,j}^{\dagger} e^{i\omega_z t})\nonumber\\
	&+& \Delta k_z z_{j}^{(0)},
\end{eqnarray}
and the Lamb-Dicke parameters are defined as
\begin{equation}
	\eta_{\beta}=\frac{\Delta k_{\beta}}{\sqrt{2m\omega_{\beta}}},~\beta=x,z.
\end{equation}

Typical distances for ions in Paul traps are $d_0$ = 10 $\mu$m. For $N = 20$ and $N = 50$ ions, and considering $^9$Be$^+$ ions, this corresponds to axial trapping frequencies $\omega_z = $ 192 and $\omega_z = $ 94 (2$\pi$) kHz, as shown by a simple calculation of the ions equilibrium positions. For an optical laser $|\Delta \mathbf{k}| \approx 2\pi / 320 $ nm\textsuperscript{-1} \cite{Leibfried03rmp}.
Let us consider now the value $\Delta k_x = 2 \pi /(3 d_0)$ used in the calculations in the main text. With the typical value for $|\Delta {\bf k}|$ above, this can be achieved by using an angle $\theta$ between the incident optical force and the perpendicular to the ion chain $\theta \approx$ 0.6 degrees, such that $\Delta k_x = \cos(\theta) |\Delta {\bf k}|$, and $\Delta k_z = \sin(\theta) |\Delta {\bf k}|$. With those values, and considering $\omega_x =$ 5 (2$\pi$ MHz)  we get Lamb-Dicke parameters $\eta_x \approx  0.21$, and $\eta_z = 0.011$ (for $N = 20$, $\omega_z = 192$ (2$\pi$) kHz) and $\eta_z = 0.016$ (for $N = 40$, $\omega_z = 98$ (2$\pi$) kHz). Since condition $\eta_x, \eta_z \ll 1$ is satisfied, we can express the spin-dependent force (\ref{spin.phonon}) like
\begin{eqnarray}
	&H_{\rm sph}&(t)=\frac{\Omega_{\rm ac}}{2} \sum_j \frac{\sigma_j^z}{2} (1+i\eta_x(a_{x,j} e^{-i\omega_x t}+a_{x,j}^{\dagger} e^{i\omega_x t})\nonumber\\ 
	&+& i\eta_z(a_{z,j} e^{-i\omega_z t}+a_{z,j}^{\dagger} e^{i\omega_z t}) ) e^{i \Delta k_z z_{j}^{(0)} }e^{-i\omega_{\rm L} t}  +{\rm H.c.},
\end{eqnarray}
where finally the lasers are made resonant with the couplings to the transversal phonons 
$\omega_{\rm L} \approx \omega_x - \delta$, with $\delta \ll \omega_x$. If we dismiss the fast rotating terms such as the carrier resonance and the interactions with the longitudinal direction (see below for a justification), we arrive at the time dependent spin-phonon Hamiltonian in the interaction picture,
\begin{equation}
	H_{\rm sph} (t)=g\sum_j \sigma_j^z \left(a_j^{\dagger} e^{i(\Delta k_z z_j^{(0)}-\delta t)}+a_j e^{-i(\Delta k_z z_j^{(0)} - \delta t)}\right),
\end{equation}
with effective spin-phonon coupling $g$ (we dropped the $x$ in the boson operators because we are dealing exclusively with the transversal oscillations). Note that in the main text we just drop the subindex $z$ in $\Delta k_z$ from $H_{\rm sph}$, and the $x$ in the radial phonon operators. 

Neglecting the spin-phonon coupling to the axial displacements, $\delta z_j$, has to be justified with care. Even though $\eta_z \ll \eta_x$, still a residual coupling to the axial modes can spoil an implementation of our proposal, since those are more difficult to cool to the ground state. However, with the range of parameters that we are choosing, the normal axial modes are not resonant with the radial ones, and the spin-dependent force can be slightly off-resonant with respect to the radial normal modes, while far detuned from the axial normal modes. For example, for $N = 20$, ($\omega_z = 192$ (2$\pi$) kHz), and $N = 50$ ($\omega_z = 0.094$ (2$\pi$) kHz), the maximum axial normal mode energy is $\omega_{z,{\rm max}} = 2.29$ (2$\pi$ MHz) and $\omega_{z,{\rm max}} = 2.5$ (2$\pi$) MHz, respectively, off-resonant with the radial normal modes, centered around $\omega_x = 5$ (2$\pi$) MHz.

\section{Effective spin-spin interaction in the long chain limit}
\label{App.decay}

In this section we show how to derive an approximate analytical expression for $J_{j,l}$ in the limit of $N\gg 1$. This can be done considering the homogeneous chain with PBC conditions. For simplicity, we also assume that $N$ is even from now on. The ions are equally spaced by a distance $d_0$, so we write $|z^{(0)}_j-z^{(0)}_l|=d_0 |j-l|$. Phonon frequencies no longer depend on the ion index $j$, and fulfill $\omega_j \simeq \omega_x,\,\forall j$, where $\omega_x$ is the trap transversal frequency. The hopping strengths $t_{j,l}$ transform into cyclic functions of $|j-l|$, i.e. $t_{j,l} = t_{\rm C} \, F_{j,l}$, where
\begin{equation}
F_{j,l} = \left\{
  \begin{array}{cl}
	0 & \text{if $j = l$,} \\
	\displaystyle \frac{1}{|j-l|^3} & \text{if $0<|j-l|\leq \displaystyle \frac{N}{2} $,}\\
	\displaystyle \frac{1}{N-|(j-l)|^3} & \text{if $\displaystyle \frac{N}{2}  <|j-l|\leq {N}$.}\\
  \end{array}
\right.
\end{equation}

Accordingly, the phonon Hamiltonian (\ref{H.ph}) reads $H_{\rm ph}(\omega_x,\{t_{j,l}\}) = \sum_{j,l} A_{j,l} a^{\dagger}_j a_l$, with
\begin{equation}
A_{j,l} = {\omega}_x \delta_{j,l}+\frac{1}{2} t_{\rm C} F_{j,l}.
\end{equation}
Every row of this matrix is a cyclically shifted version of the previous upper row. Thus, it is diagonal in the basis of eigenstates ${ M}_{j,n}= e^{i\frac{2\pi n}{N}j}/\sqrt{N},~n=0 ,\ldots,N-1$. Furthermore, the dispersion relation of the normal modes, $\omega_n=\sum_m\sum_{j,l}A_{j,l}M_{j,n}{M}_{l,m}^*$, is just the discrete Fourier transform of any of the rows of $F_{j,l}$. We parametrize these as $F_q= F_{|j-l|}$, $q=0,\ldots,N/2 $, so as
\begin{equation}
\omega_n=\omega_x +\frac{t_{\rm C}}{2} \left(F_{N/2} (-1)^n +  \sum _{q=0}^{N/2 -1}2\cos(\frac{2\pi q n}{N})F_{q}\right).
\end{equation}
We note that this expression, as a function of $n$, has a minimum for $n=N/2$. This mode describes ions oscillating in a zigzag structure along the trap axis (see Fig. \ref{scheme}, blue line). It has the lowest energy among the collective excitations, because the vibrational modes $M_{j,n}$ stem from the radial off-equilibrium displacements of the ions, for which $t_{\rm C}>0$, meaning that spatial orders where the ions are as far apart from each other as possible, are energetically preferred.
\begin{figure}[h!]
		\includegraphics[width=3.5in]{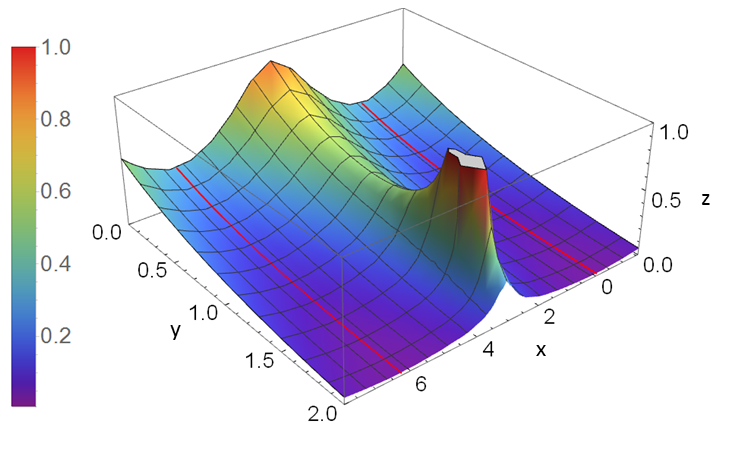} 
        \caption{(color online) Plot of $z=|f(z)|$, where $f(z),\,z=x+iy$, is the integrand of the last integral in (\ref{contour.integral}). We set $\delta_{N/2}=2,t_{\rm C}=1,g=1$ and $|j-l|=1$. The  lines along $x=0,2\pi$ depict the trajectories of fastest descend, which go exponentially to zero. There is a pole at $z_+ \simeq \pi + i1.5$.}
        \label{analytic.landscape}
\end{figure}

The former frequencies are shifted once we move into a rotating frame with the laser detunings, so as $\omega_n - \omega_{\rm L}=\delta_n$. In order to understand the decay of $J_{j,l}$ it is convenient to work out its continuum limit, e.g.
\begin{equation}
J_{j,l}^{\Delta k=0} = -\sum_{n=0}^{N-1} \frac{g^2}{\delta_n} \frac{1}{N}e^{i \frac{2\pi n}{N}(j-l)} \overset{N\to \infty}{\sim}-\frac{1}{2\pi} \int_{0}^{2\pi} \frac{g^2 e^{ix|j-l|}}{\delta(x)}dx,
\label{integral.of.coupling}
\end{equation}
where the dispersion relation $\delta(x)$ is now the real valued function
\begin{equation}
\delta(x) = \delta_{x} + t_{\rm C}\sum_{k=1}^{\infty}\frac{\cos(kx)}{k^3}, \quad \delta_x = \omega_x - \omega_{\rm L}.
\label{real.dispersion.relation}
\end{equation}

In accordance with the previous discussion, we find that $\delta(\pi) = \delta_x - 3/4 t_{\rm C} \zeta(3)$ is the minimum value of $\delta(x)$ on the interval $[0,2\pi]$ ($\zeta(3)\simeq 1.2021$, where $\zeta$ is the Riemann Zeta function). We will refer to it as $\delta_{N/2}=\delta_{N/2}(\delta_x, t_{\rm C})$. It is positive if $\delta_{x}>3/4t_{\rm C}\zeta(3)$. We consider this scenario of $\delta_{N/2}>0$ in the following. Negative phonon energies $\delta_{N/2}<0$, on the contrary, lead to an unbounded (from below) energy spectrum for $H_{\rm ph}(\{\delta_n\})$, as well as the appearance of zeros in the denominator of (\ref{integral.of.coupling}) along its domain of integration.
\begin{figure*}[t!]
        \begin{subfigure} []
        {\includegraphics[width=3in]{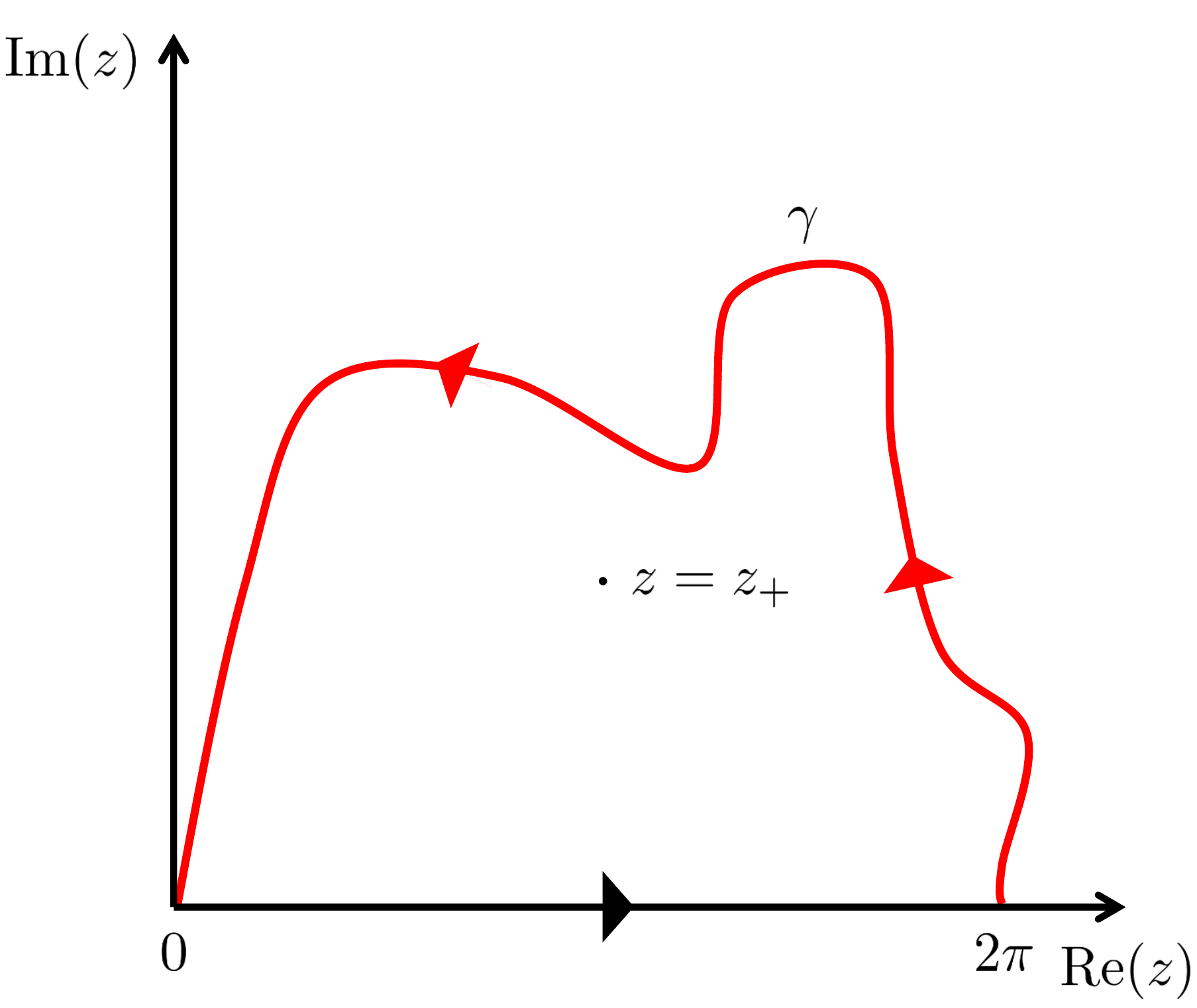} \label{exact.contour}}
        \end{subfigure}
        \centering
        ~
        \begin{subfigure} []
        {\includegraphics[width=3.1in]{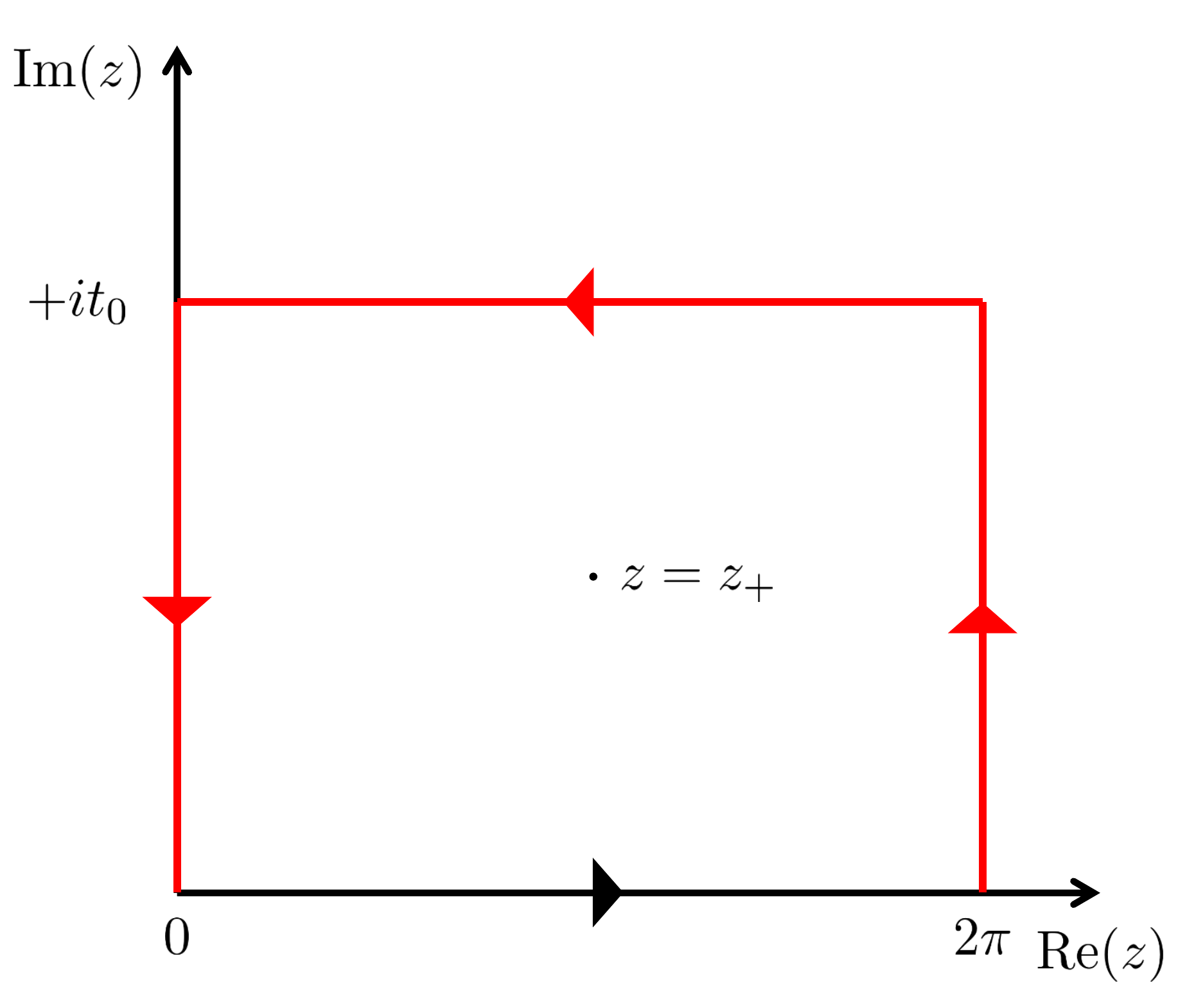} \label{steepest.contour}}
        \end{subfigure}
        \caption{(color online)  
        (a) Closed contour consisting of the segment along the real line from $z=0$ to $z=2\pi$ and the curve in the upper part of the complex plane $\gamma$. The black dot marks the position of $z_+$, the approximate root of $\delta(z)$ with $\operatorname{Im}(z_+)>0$.
        (b) Effective contours for the evaluation of the second term in (\ref{contour.integral}). These are the next order contributions after the residue in the pole $z_+$. Along these directions the integrand shrinks exponentially fast. Because of this we consider also that the non-exponential part of the integrand is well described by its series expansion close to the real axis.}
\end{figure*}

For the evaluation of (\ref{integral.of.coupling}) we are going to rely on its extension to the complex plane. We start by noticing that the complex-valued function 
\begin{equation}
\delta(z)=\delta_x + \frac{t_{\rm C}}{2} (\Li_3(e^{-iz}) + \Li_3(e^{iz})),\,z=x+iy,
\label{dispersion.relation}
\end{equation}
coincides with the dispersion relation (\ref{real.dispersion.relation}) when $y=0$. In this expression,
\begin{equation}
\Li_{3}(z)= \sum_{k=1}^{\infty} \frac{z^k}{k^3},\,|z|\leq 1
\end{equation}
is the polylogarithm function, which admits this series representation on the unit circle centered around the origin of the complex plane. Now we can use this function, together with the Residue Theorem (cf. \cite{Arfken.book}) to write
\begin{align}
&-\frac{1}{2\pi} \int_{0}^{2\pi} \frac{g^2 e^{ix|j-l|}}{\delta_x + t_{\rm C}\sum_{k=1}^{\infty}\frac{\cos(kx)}{k^3}}dx \nonumber\\
&= 2\pi i  \operatorname{Res}(\frac{-g^2 e^{iz|j-l|}}{2\pi\delta(z)},\,z_{+})+\frac{1}{2\pi} \int_{\gamma} \frac{g^2 e^{iz|j-l|}}{\delta(z)}dz.
\label{contour.integral}
\end{align}

The integrand on the complex plane is visualized in Fig. \ref{analytic.landscape}. We denote the zero of $\delta(z)$ within the contour depicted in Fig. \ref{exact.contour} by $z_+$. We choose this contour as the segment $[0,2\pi]$ along the real axis, and the curve $\gamma$ in the upper complex plane. At this time, $\gamma$ is completely arbitrary, apart from enclosing $z_+$.

Let us focus first on computing the residue term in (\ref{contour.integral}). As already noted, the function $\delta(z)$ has no zeros on the real line as long as $\delta_{x} > 3/4t_{\rm C} \zeta(3)$. In the complex plane, however, there are two zeros of $\delta(z)$ along the line $x=\pi$, which are symmetric with respect to the real axis. To give an analytical estimation, we further assume that they are of the form $z = \pi + iy_{\pm}$, with $|y_{\pm}|\ll 1$. Therefore, the function (\ref{dispersion.relation}) is approximated by the first non-trivial term in its Taylor series,
\begin{align}
\delta(\pi + iy_{\pm}) &\simeq \delta_x + t_{\rm C} \sum_{k=1}^{\infty} \frac{(-1)^k}{k^3}(1 - \frac{k^2y_{\pm}^2}{2}) \nonumber\\
&= \delta_x -\frac{3}{4} \zeta(3) t_{\rm C} - y_{\pm}^2 \log(2)\frac{t_{\rm C}}{2}.
\label{delta.approx}
\end{align}
Thus, $\delta(\pi+iy_{\pm}) \simeq 0$ is attained for
\begin{equation}
y_{\pm} = \pm\sqrt{\frac{\delta_x -3/4 \zeta(3) t_{\rm C}}{t_{\rm C} \log{2}/2}},
\end{equation}
Equivalently, we can write $\delta(z_{\pm }) \simeq 0 \Rightarrow z_{\pm}=\pi + i y_{\pm}$. Since we are only interested in the residue of $\delta(z)$ at the pole, we further assume that $\delta(z)\simeq (z-z_+)(z-z_-)$, which makes the calculation straightforward, and yields
\begin{equation}
2\pi i \operatorname{Res}(\frac{-g^2 e^{iz|j-l|}}{2\pi\delta(z)},\,z_+) \simeq  -(-1)^{|j-l|}\frac{g^2\xi}{t_{\rm C}\log(2)}e^{-\frac{|j-l|}{\xi}},
\end{equation}
where $\xi = |y_+|^{-1}$. We recall that approximating $\delta(z)$ in this way is justified only when the pole $z_+$ is very close to the real axis. Therefore, we expect a better agreement for bigger $\xi$. This exponential decay is the continuous version of the laser being resonant with one or several normal modes, whose effective spread upon the chain is $\xi$.

The second term in (\ref{contour.integral}) can be estimated from the contributions of the integrand upon its fastest descent directions. This amounts to a particular choice of $\gamma$, with the only constraints of having its endpoints on the real axis, and enclosing the point $z_{+}$. The directions for which $|e^{iz|j-l|}|/|\delta(z)|$ shows the largest changes are lines perpendicular to the real axis, starting at $z=0$ and $z=2\pi$, and running towards the upper part of the complex plane. This is so because of the exponential decrease of the integrand along them. Therefore, we choose $\gamma$ as the segments parametrized by $z_1(t)=it$ and $z_2(t)=2\pi +it$ for $t\in[0,t_0],t_0>0$, and a line joining them at their endpoints (cf. Fig. \ref{steepest.contour}). The contribution of the latter is negligible as the integrand is exponentially shrunk away from the real axis. Therefore
\begin{align}
&\frac{1}{2\pi} \int_{\gamma} \frac{g^2 e^{iz|j-l|}}{\delta(z)}dz \simeq \frac{g^2}{2\pi}\left(\int_{t_0}^0 \frac{e^{-t|j-l|}}{\delta(it)}idt +\int_{0}^{t_0} \frac{e^{-t|j-l|}}{\delta(2\pi+it)}idt\right)=\nonumber\\
&\frac{g^2}{2\pi} \int_{0}^{t_0}dt\,i e^{-t|j-l|}\left(\frac{1}{\delta(2\pi+it)}-\frac{1}{\delta(it)}\right)=\nonumber\\
&\frac{g^2}{2\pi} \int_{0}^{t_0}dt\,i e^{-t|j-l|}(-2i)\operatorname{Im}\left(\frac{1}{\delta(it)}\right)\simeq \nonumber\\
&\frac{g^2t_{\rm C}}{4(\delta_x+t_{\rm C}\zeta(3))^2} \int_{0}^{\infty} t^2e^{-t|j-l|} dt.
\label{cubic.integral}
\end{align}

In this derivation, we have relied on the fact that $\delta(it)$ and $\delta(2\pi + it)$ belong to different branches of the polylogarithm, for which the imaginary part changes its sign. Also, we approximate $1/\delta(it)$ by its series expansion, to second order, for $t\to 0^+$, since these are the only values not exponentially eliminated by $e^{-t|j-l|}$. We have assumed as well that no further error is included if we make $t_0\to \infty$. Performing the previous integral, we arrive at
\begin{equation}
\frac{1}{2\pi} \int_{\gamma} \frac{g^2 e^{iz|j-l|}}{\delta(z)}dz \simeq \frac{g^2 t_{\rm C}}{2(\delta_x+t_{\rm C}\zeta(3))^2} \frac{1}{|j-l|^3},\,|j-l|\gg 1.
\end{equation}

We conclude that the effective coupling of the Ising interaction stemming from the Coulomb repulsion has two main contributions, e.g.
\begin{equation}
J^{\Delta k = 0}_{j,l} \simeq - (-1)^{j-l} J_{\rm exp} e^{- \left| j - l \right|/\xi}
+ \frac{J_{\rm dip}}{\left| j - l \right|^3}_{j \neq l},
\label{continuum.approx}
\end{equation}
where 
\begin{equation}
J_{\rm exp} = \frac{g^2\xi}{t_{\rm C}\log(2)},\, J_{\rm dip}=\frac{g^2t_{\rm C}}{2(\delta_{N/2}+7/4\,t_{\rm C}\zeta(3))^2}.
\end{equation}

The interplay of the exponential and dipolar decays is dictated by the magnitude of the typical length $\xi$, as given by
\begin{equation}
\xi = \sqrt{{\log{2}/}{2}}\sqrt{{t_{\rm C}/}{\delta_{N/2}}}.
\end{equation}

If $\delta_{N/2}\gg t_{\rm C}$, the exponential behavior is overtaken by the dipolar contribution. This effect is usually attributed to a destructive interaction between normal modes, since all of them are equally off-resonant with the laser \cite{Schneider12rpp}. If $\delta_{N/2}\ \sim t_{\rm C}$, the laser frequency lies within the motional spectrum. This frequency will be close to one or several $\delta_n$. Therefore, $\xi$ is a measure of the extent of these motional wave functions.

\bibliography{biblio}

\end{document}